# Phenomenological Arrhenius Analyses in Plasmon-Enhanced Catalysis

*Prashant K. Jain*[1,2,3,4*]

[1]Department of Chemistry, [2]Materials Research Laboratory, [3]Department of Physics, and [4]Beckman Institute for Advanced Science and Technology, University of Illinois at Urbana-Champaign, Urbana, IL 61801, USA

*Corresponding Author E-mail: jain@illinois.edu

In the recently published article: "Thermal effects–an alternative mechanism for plasmon-assisted photocatalysis", Dubi *et al.*[1] argue that the results of multiple works on plasmon-excited-induced bond dissociation reactions can be explained by a purely photothermal enhancement of the reaction rates; no non-thermal effects are required to explain the enhanced rates resulting from plasmonic excitation. Their argument rests on a reproduction of the reaction rate data using an Arrhenius expression with a light-intensity-dependent local temperature at the surface of the nanoparticles.

Dubi *et al.*'s straightforward analysis may have general appeal for explaining rate enhancements in bond dissociation reactions observed under plasmonic excitation of metal nanostructures without invoking hot electron contributions. But there is one caveat that deserves recognition when undertaking such an analysis. As shown below, under certain common scenarios, it is practically impossible to distinguish between a photochemical (non-thermal) effect of light excitation and a purely photothermal one using a phenomenological Arrhenius fitting of the data alone.

As per the Arrhenius equation, the rate of a reaction depends on the set temperature $T_s$ as:

$$R = R_0 \cdot \exp\left(\frac{-E_\mathrm{a}}{k_B T_s}\right) \quad (1)$$

where $R_0$ is a constant for a given reaction and reaction conditions and $E_a$ is the apparent activation energy barrier for the reaction. As an aside, one should note that unlike the Eyring equation, which is preferred for non-gas-phase reaction kinetics involving a vibrational reaction co-ordinate, the pre-exponential factor in the Arrhenius equation is assumed to have a negligible temperature dependence.

A photochemical explanation of plasmon-enhanced catalysis is that the apparent activation energy $E_\mathrm{a}$ is lower under plasmonic excitation as compared to its value, $E_\mathrm{a}^\mathrm{dark}$, in the dark. Thus, as per eq. (1), at a fixed temperature $T_s$, $R$ will be higher under light excitation. In fact, the measured apparent activation barrier has been found to be dependent on the light intensity $I$. For the sake of the following argument, let us assume that the decrease in $E_a$ is linearly dependent on the light intensity:

$$E_a = E_a^{dark} - B.I \quad (2)$$

where $B$ is a proportionality constant with units of eV.cm$^2$.W$^{-1}$ when $E_a$ is expressed in units of eV and $I$ in units of W.cm$^{-2}$. Note that $B$ is expected to be wavelength-dependent. Eq. (2) can be written alternatively as:

$$E_a = E_a^{dark}(1 - b.I) \quad (3)$$

where $b$ is simply $B/E_a^{dark}$ and has units of cm$^2$.W$^{-1}$. From eqns. (1) and (3):

$$R = A.\exp\left(\frac{-E_a^{dark}}{k_B T_s}(1 - bI)\right) \quad (4)$$

Using a Taylor's expansion around $I = 0$ (dark condition),

$$\frac{1}{1-bI} = 1 + bI + (bI)^2 + \cdots \quad (5)$$

For the light-intensity regime ($I \ll 1/b$), the higher order terms can be neglected, so one gets from eqs. (4) and (5):

$$R = R_0.\exp\left(\frac{-E_a^{dark}}{k_B T_s(1+bI)}\right) \quad (6)$$

Thus, if one simply uses an Arrhenius analysis of the reaction rate, the reaction appears to be carried out at a hypothetical temperature that is higher than the actual temperature $T_s$ by an amount proportional to the light intensity $I$:

$$T_{dummy} = T_s(1 + bI) \quad (7)$$

where this hypothetical temperature is referred to as $T_{dummy}$. Eq. (7) is equivalently expressed as:

$$T_{dummy} = T_s + aI \quad (8)$$

where $a = b.T_s$ is the photothermal conversion coefficient with units of K.cm$^2$.W$^{-1}$. Eq. (8) is identical to the expression used by Sivan *et al.* in their argument in favor of a purely photothermal effect. In other words, it would appear as if plasmonic excitation led to an increase in the temperature, but led to no change in the apparent activation barrier. Effectively, in a phenomenological Arrhenius analysis, the photochemical (non-thermal) effect of plasmonic excitation on the reaction is simply masked as a temperature increase.

Thus, as shown in Figure 1, an Arrhenius analysis with $a$ as an adjustable fit parameter may be futile for practically distinguishing the photochemical action of plasmonic excitation, (i.e., a rate enhancement caused by a decrease in the activation barrier) from a purely photothermal effect (i.e., a rate enhancement caused by an increase in the surface temperature). Under such a scenario, for distinguishing these effects, it is necessary to have precise knowledge and/or control over the temperature at the surface of the nanoparticles, as correctly argued by Dubi *et al.*,[1] but also acknowledged by practitioners[2–4] in the field. It is well appreciated that the localized inhomogeneous nature of photothermal heating results in a temperature gradient extending out from the surface of the nanoparticles to the bulk of the medium. These gradients are small in magnitude under conditions where the heat dissipation rate can keep up with the energy deposition rate. However, in systems where heat transfer rates are limiting, significant non-uniformities in temperature and thermal bottlenecks can arise. Such cases necessitate spatially precise temperature-probing localized to the nanoparticle surface.

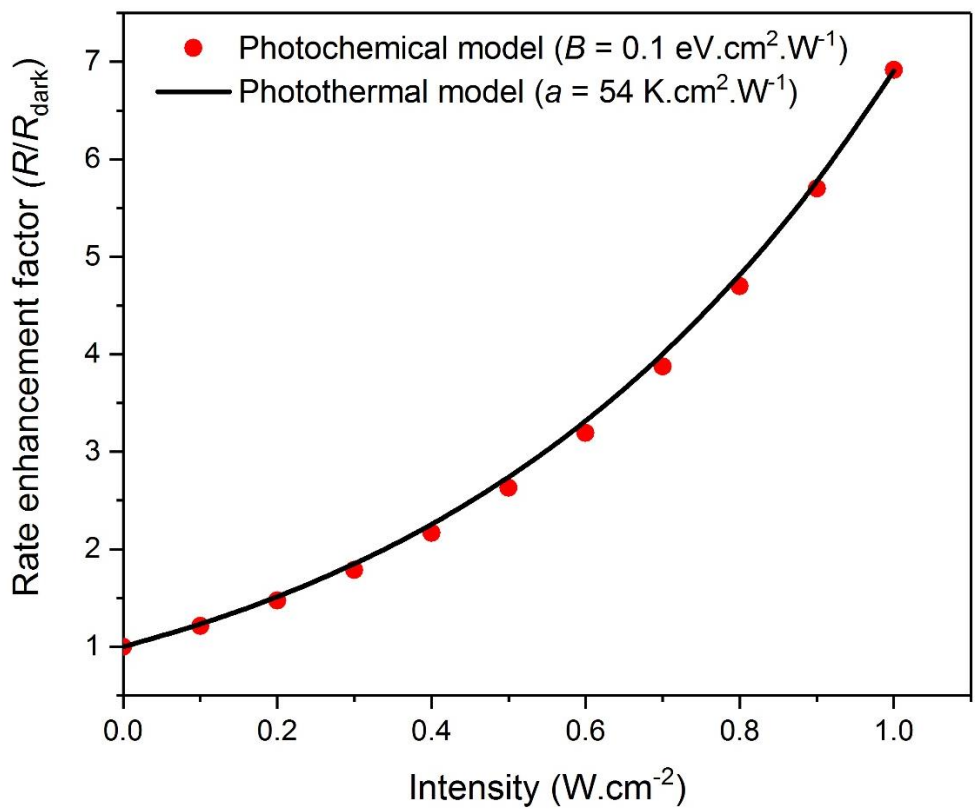

**Figure 1.** The reaction rate under plasmonic excitation, $R$, relative to that in the dark, $R_{dark}$, is plotted as a function of light intensity for i) the photochemical case (red dots), where the activation barrier is decreased by plasmonic excitation (eqs. (1) and (2) with $B = 0.1$ eV.cm$^2$.W$^{-1}$) while the temperature is kept fixed and ii) the purely photothermal model (black line), where the temperature is increased by plasmonic excitation (eqs. (1) and 8) with $a = 54$ K.cm$^2$.W$^{-1}$) but the activation barrier remains unchanged. In both cases, $E_a^{dark} = 1.21$ eV and $T_s = 600$ K. The two models yield trends that are practically indistinguishable.

**Acknowledgements**. This material is based upon work supported by the National Science Foundation under Grant (NSF CHE-1455011). The author thanks Varun Mohan for proof-reading. The author has no competing interests.